\documentclass[]{aa}
\usepackage{graphicx}
\begin{document}

\title{On the difference between type E and A OH/IR stars.}

\author{J.H. He and P.S. Chen}

       \offprints{J.H. He, \\ mailhejh@yahoo.com.cn}

       \institute{\centering{Yunnan Observatory, Chinese Academy of Sciences, Kunming 650011, China\\
       United Laboratory of Optical Astronomy, CAS\\
       National Astronomical Observatories, China\\
       Email: mailhejh@yahoo.com.cn}
                 }
          \date{Received / Accepted }

       \abstract{
The observed SEDs of a sample of 60 OH/IR stars are fitted using a radiative
transfer model of a dusty envelope. Among the whole sample, 21 stars
have reliable phase-lag distances while the others have less accurate distances. 
$L_{*}-P$,$\dot{M}-P$ and $\dot{M}-L_{*}$ relations have been plotted for these
stars. It is found that type E (with emission feature at $10\mu$) and
type A (with absorption feature at $10\mu$)  OH/IR stars have different $L_*-P$
and $\dot M-L_{*}$ relations while both of them follow a single $\dot M-P$ relation. The type E
stars are proven to be located in the area without large scale dense
interstellar medium while the type A stars are located probably
in dense interstellar medium. It is argued here that this may indicate the two
types of OH/IR stars have different chemical composition or zero age main sequence mass
and so evolve in
different ways. This conclusion has reinforced the argument by \cite{che01}
who reached a similar conclusion from the galactic distribution of about 1000
OH/IR stars with the IRAS low-resolution spectra (LRS). 

}

	\titlerunning{OH/IR stars of type E and A}
	\authorrunning{He \& Chen}

       \maketitle

\section{Introduction}

Recently \cite{che01} have published a catalogue of OH/IR stars with
IRAS LRS spectra. They also argued that many type E OH/IR stars which are located
at high galactic latitude will never evolve into type A OH/IR stars
. In order to examine the differences
between the two groups of OH/IR stars (type E and A), it is tried in this paper
to investigate the luminosity-period relation and mass loss rate-period
relation of the two groups by radiative transfer modeling.  The observed data 
and the radiative transfer model used are described in sections 2 and
3, and a detailed discussion of the derived quantities and conclusions can be 
found in section 4.

\section{Data}

A sample of 60 OH/IR stars with known distances is collected from the literature; 36 of
them belong to type E, 18 belong to type A. For the purpose of comparison, 3 
type H stars (red continuum), 2 type U stars (unusual spectrum) and 1 type F 
star (featureless) are also included. The observational SED (Spectral Energy
Distribution) of each star consists of three parts: near infrared (NIR)
photometry taken from ground based observations, IRAS far infrared (FIR)
photometry and IRAS LRS spectrum. The LRS spectra are from the
on-line IRAS Data Analysis Facility of the University of Calgary.
The FIR fluxes are obtained from the CD-edition of
the published IRAS PSC catalogue and only those fluxes with good quality ($Q\geq
2$) are used. The NIR photometry data of 21 OH/IR stars are
given by \cite{xio94}, and that of the remaining 39 stars are taken from
different papers. The references for NIR photometry data are listed in Table
1. The NIR data are corrected for interstellar extinction. The formula of 
\cite{her65} is used:
\begin{equation}
A_{V}=0.14\csc \left| b \right| \left[ 1-\exp \left(
-10d\sin \left| b \right| \right) \right]
\end{equation}
where b is the galactic latitude, d is the distance from the sun. We also use
the method from \cite{car89} to calibrate the data at different
wavelengths.

The periods of pulsation for 46 stars are also searched from \cite{che01}.
 Twinty one sample stars have the so called phase-lag
distances (this is the most accurate distance available for OH/IR stars up till now,
and hereafter these stars are called as the phase-lag sample, and the whole 60
 stars as whole sample), and these distances are taken from \cite{bau83},
\cite{her86} and \cite{lan90}. The distances of
the other stars are taken from different papers where they are estimated by
other methods, e.g. kinematic method, using $L_{OH}-R^{2}$ relation and
so on, and so they are not very accurate (error can be larger than $100\%$).
The references for distance are
listed also in Table 1. For the phase-lag sample, the periods, expansion
velocities and phase-lag distances are shown in Table 2.

\section{Modeling}

\subsection{Model description}

The radiative transfer model of a spherical dusty envelope is used to fit the
observed SEDs. The model used is originally compiled by \cite{szc97}
. This code is able to calculate the emerging flux spectrum of a
spherical shell observed at a specified distance when the luminosity ($L_{*}$) and
effective temperature ($T_{eff}$) of the central star and dust species, dust
condensation temperature ($T_{cond}$), dust density distribution ($\rho _{d}\sim 
r^{-n}$), grain size distribution and the geometrical 
sizes of the envelope are given. The original code is slightly modified so as to make it 
possible to vary only the following five quantities when fitting SEDs: $T_{eff}$,$T_{cond}$,$\tau _{V}$,$n$ and
$Y_{max}$, while the other quantities are fixed at their typical values:
astronomical silicate grains whose electric permittivity and magnetic
permeability are taken from \cite{dra84} and \cite{dra85}; grain size
distribution of interstellar dust as $n\left( a \right) \sim a^{-3.5}$
with the grain size range from $a_{min}=0.005\mu$ to $a_{max}=0.25\mu$\cite{mat77}
. Here $\tau _{V}$ is the total optical thickness at $550nm$, $n$ is the index of
negative exponential density distribution law and $Y_{max}$ is the relative
geometrical thickness $R_{max}/R_{min}$. The extinction coefficient and the albedo
of dust grains are calculated from Mie theory. In order to fit the observed
SEDs, 2475 models have been calculated by varying $T_{eff}$ from $1300$K to
$3800$K,$T_{cond}$ from $800$K to $1000$K,$\tau _{V}$ from $0.1$ to $80$,$n$ form $1.4$ to $2.6$ and
$Y_{max}$ from $10^{3}$ to $10^{5}$.

\subsection{Model fitting}

The models calculated above are used to fit the observed SEDs simply by
weighted linear interpolation. The weights are given larger for LRS spectrum
and FIR data and smaller for the NIR data because the NIR fluxes are varied for most sample stars.
After having derived
$T_{eff}$,$\tau _{V}$,$n$,$Y_{max}$ and the best fitted model SED for each star from above, it
is easy to determine the luminosity ($L_{*}$) and gas mass loss rate ($\dot M$) (when the
outflow velocity is known) by assuming the dust to gas mass ratio to be $10^{
-3}$ for all stars. Here it should be pointed out that the mass loss rate is calculated at 
the inner border of the circumstellar shell, hence this quantity is theoretically
in the simultaneous measure with the luminosity of the center star. Some
derived quantities of the phase-lag sample stars are listed also in Table 2.

\section{Analysis on the fitted results}

Firstly the luminosity-period relation ($L_{*}-P$ relation) for the phase-lag sample of OH/IR stars 
is shown in Fig. 1. As we can see, for both type E and A 
stars, the luminosity seems to be larger for longer period, but the
luminosity for the two types seem to be going along parallel paths, that is
to say, the two $L_{*}-P$ relations do not join with each other smoothly from
that of the type E to that of the type A as we expected. But it is difficult to
draw any convincing conclusion from Fig. 1, because too few type E OH/IR
stars are included in this figure. Then the same figure for the whole sample is displayed
in Fig. 2 where many more type E stars are included. This figure
is proven to confirm the conclusion of Fig. 1. The $L_{*}-P$ relation does
follow different tracks for the type E and A OH/IR stars, although the
dispersion in the $L_{*}-P$ relation is large for both types. In both Fig. 1
and 2, type H stars are located in type A region and seem to follow the $L_{*}-P$
relation of type A stars.
\begin{figure}[]
   \centering
	 \includegraphics[width=9cm]{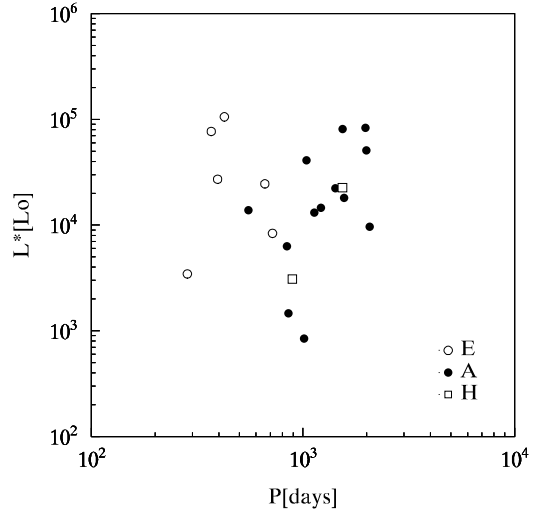}
   \caption{The $L*-P$ relation for the phase-lag sample. The luminosity shown is
obtained by modeling.}	
   \label{Fig1}
   \end{figure}
\begin{figure}[]
   \centering
	 \includegraphics[width=9cm]{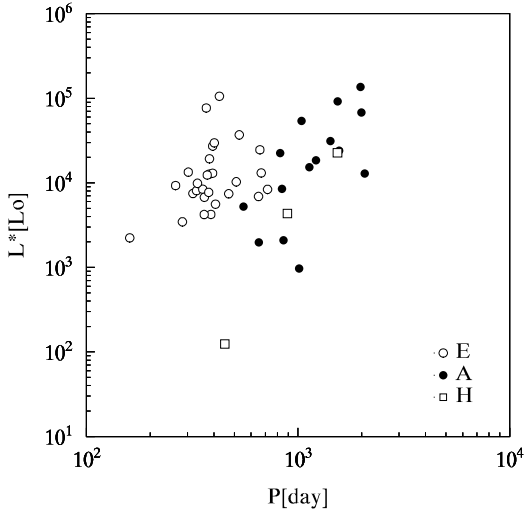}
   \caption{The $L*-P$ relation for the whole sample. The
luminosity shown is obtained by modeling.}	
   \label{Fig2}
   \end{figure}
\begin{figure}[]
   \centering
	 \includegraphics[width=9cm]{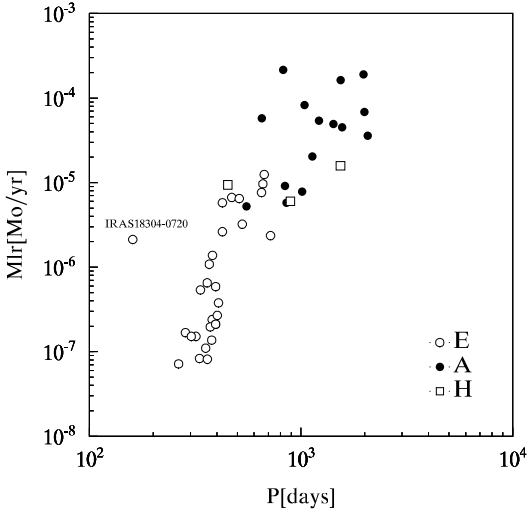}
   \caption{The $\dot M-P$ relation for the OH/IR stars with known outflow velocity of
the whole sample. The mass loss rate shown is obtained by modeling. }	
   \label{Fig3}
   \end{figure}

The $L_{*}-P$ relation of OH/IR stars and Miras had been addressed by many other authors 
\cite{whi91,ber93,blo98}, but all of them had considered only a small number 
of OH/IR stars and taken them as a single class. Here we divid the OH/IR stars into different
subclasses according to their IRAS LRS classification and more stars have been included. This 
give us the chance to see some inner structures of the $L_{*}-P$ relation for OH/IR stars, as 
described above.
 
The mass loss rate-period relation ($\dot M-P$ relation) of the whole sample is plotted
in Fig. 3. It is easy to see that the whole sample follows a
single $\dot M-P$ relation and, just as we already know, the mass loss rates increase
from type E stars with shorter period to type A stars with longer period. Similar 
$\dot M-P$ relation had ever been found before for M-type stars by some other authors, 
eg. \cite{whi94}. It
should be noted that one star is located far from the majority; it is
IRAS 18304-0720. This star is found to be seriously contaminated by cirrus
emission, and extended structures have been detected around it in all four IRAS
bands ($12,25,60,100\mu$). Hence the observed spectral energy distribution of
this star no longer represents that of a single shell.

The $\dot M-L_{*}$ relation is shown in Fig. 4. In this figure the 
type E stars seem to distribute in an area different from that of the stars of other types.
Although the luminosity of type E stars can be very large, the mass loss
rate seems always to be smaller than that of the other types with the same
luminosity.
\begin{figure}[]
   \centering
	 \includegraphics[width=9cm]{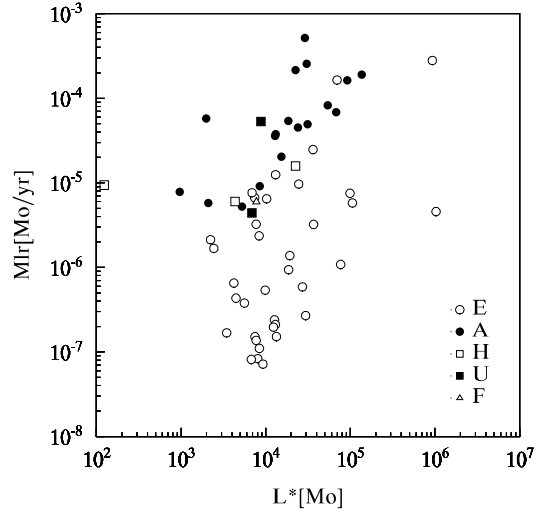}
   \caption{The $\dot M-L*$ relation for the OH/IR stars with known outflow velocity of the
whole sample. The mass loss rate and luminosity shown are obtained by
modeling. }	
   \label{Fig4}
   \end{figure}
\begin{figure}[]
   \centering
	 \includegraphics[width=9cm]{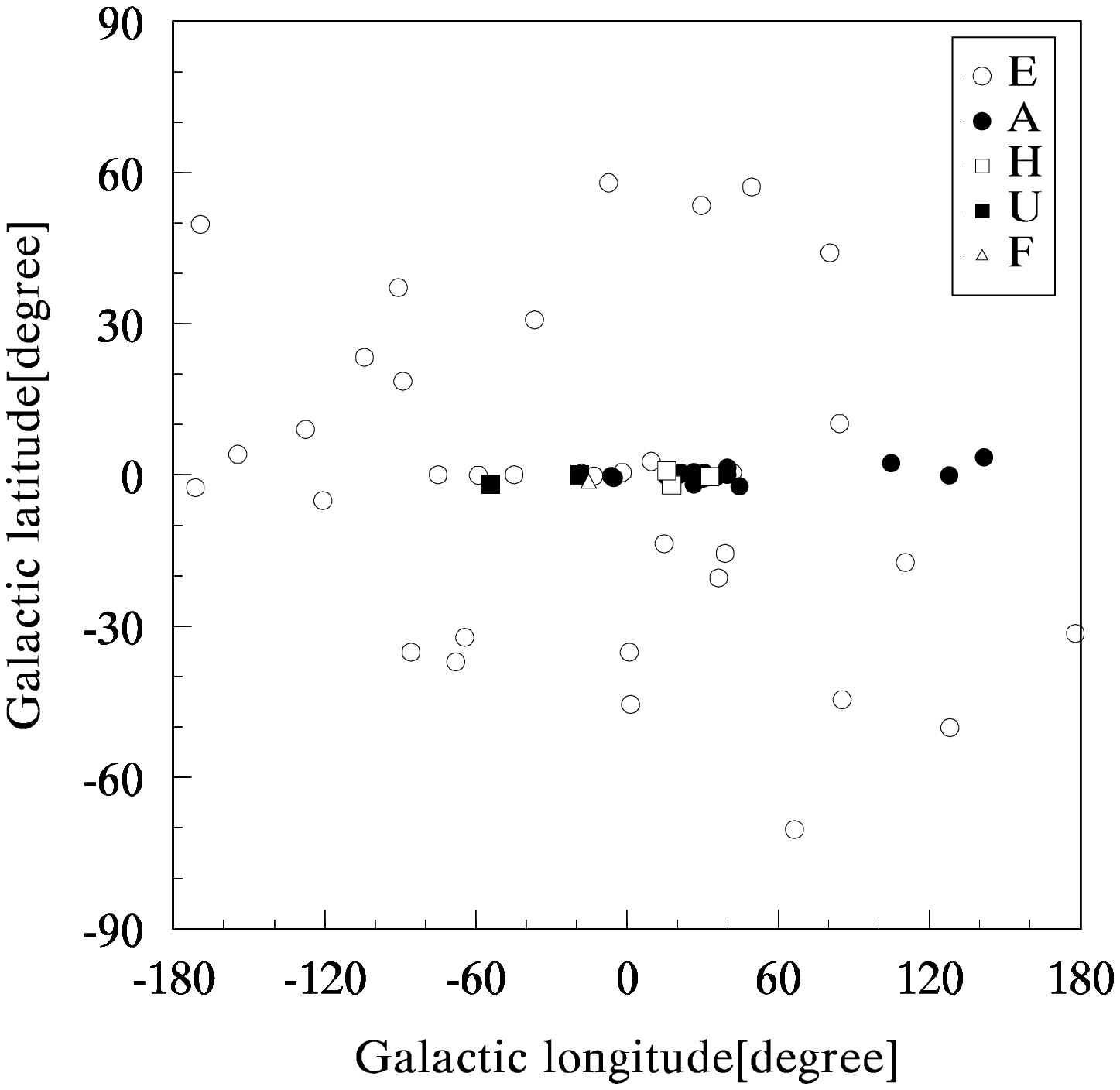}
	 \includegraphics[width=9cm]{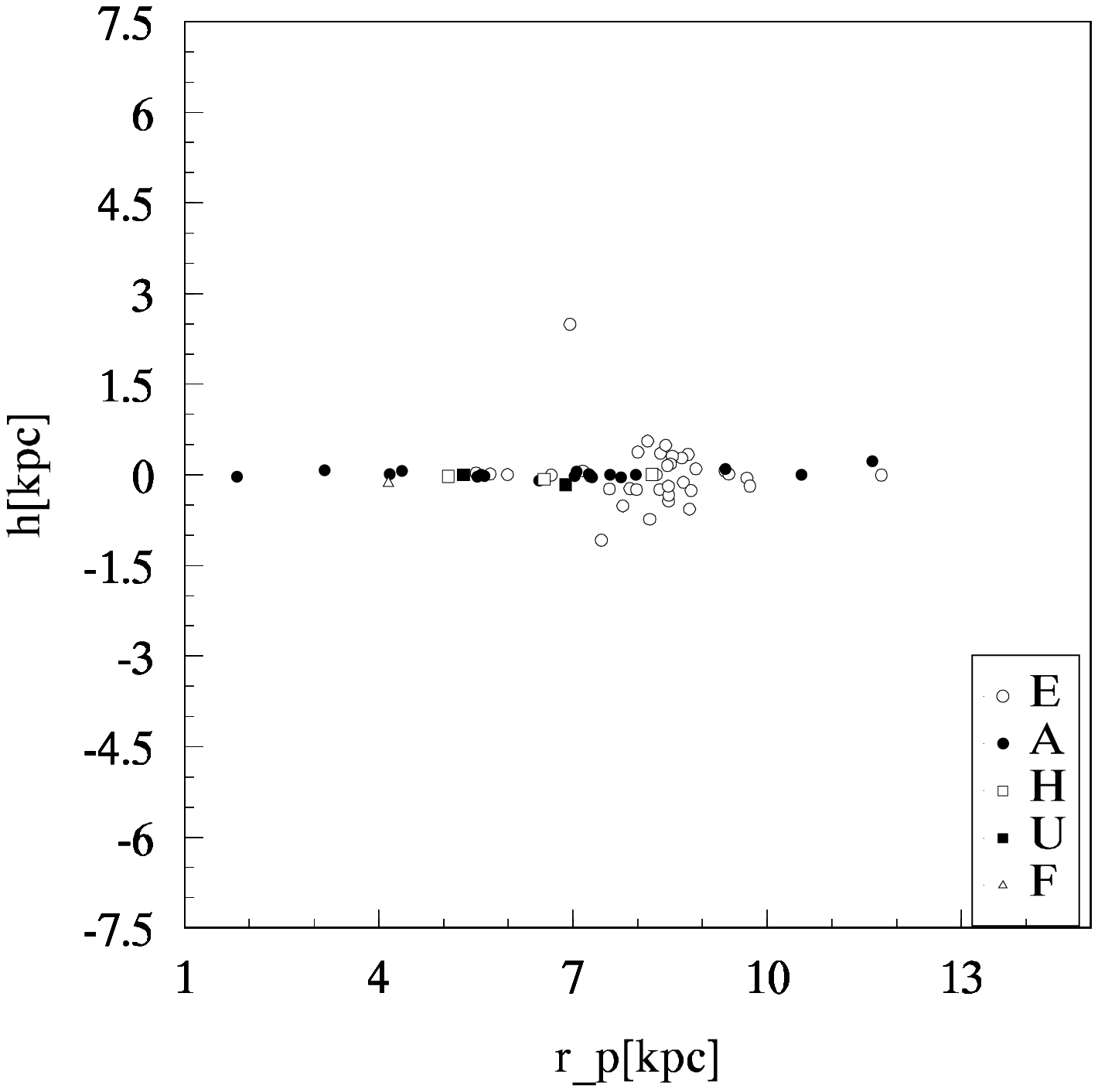}
	 \includegraphics[width=9cm]{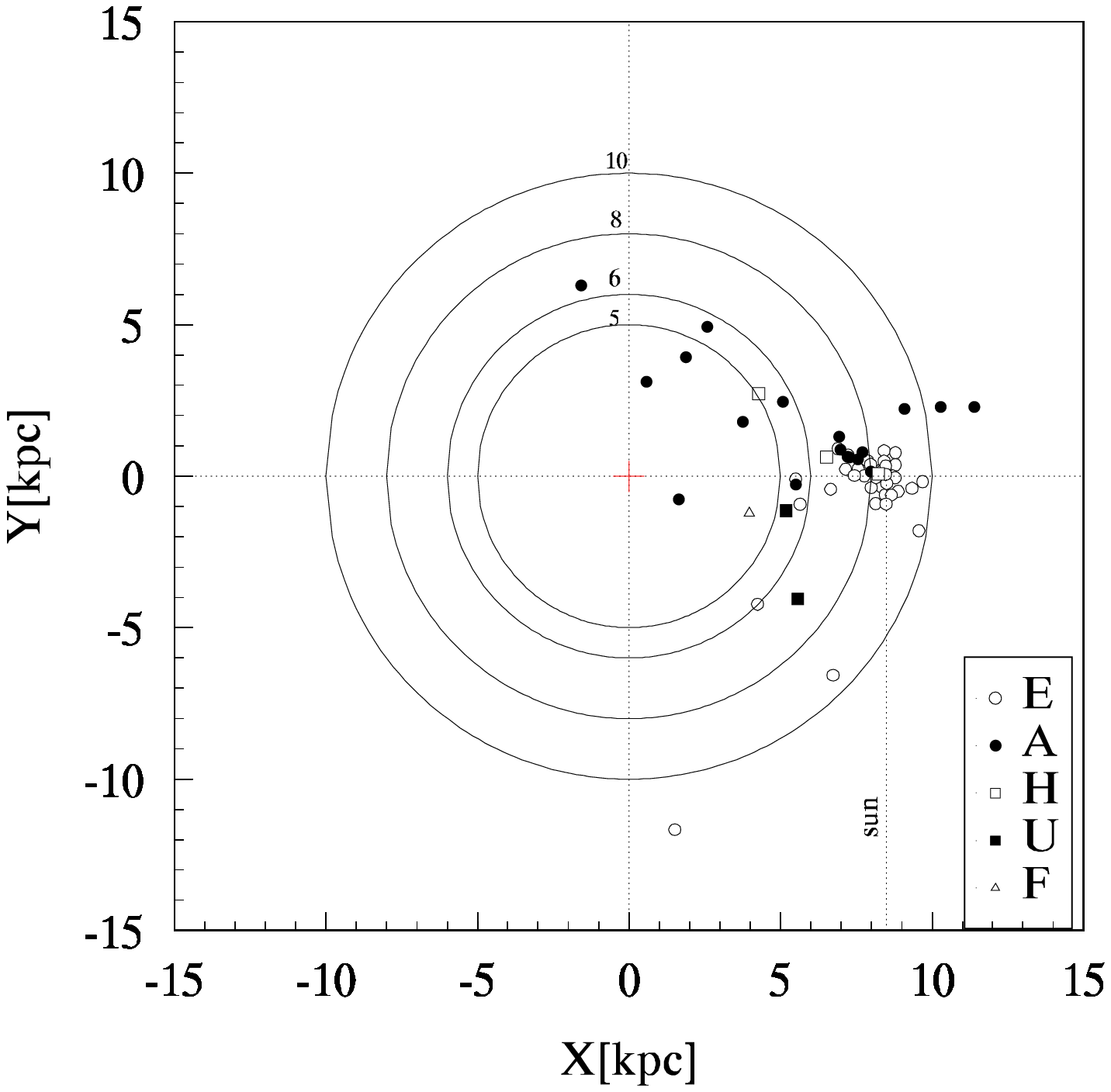}
   \caption{ a) The space distribution of the whole sample on the galactic
 coordinate. b) Distances of stars to the galactic plane and the galactic
 center. $h$: distance to galactic plane; $r\_p$: distance from the projections of
 stars on the Galactic plane to the galactic center. c) the projections of sample stars on the
 galactic plane. The galactic center is marked by ''+'', the position of the
 sun is marked out by a dotted line. Numbers beside those circles denote the
 radius of them in kpc. }	
   \label{Fig5abc}
   \end{figure}

From the discussion above,
the $L_{*}-P$ and $\dot M-L_{*}$ relation of type E and A stars seems to follow different
tracks while the $\dot M-P$ relation of all types in the whole sample follow the
same track. What do these facts mean to the evolution of OH/IR star? According to the general
evolutionary picture of AGB star, the luminosity and the mass loss rate of the AGB stars
should increase with time and both of them should be the function of the zero age main 
sequence mass ($M_{zams}$) and chemical abundance. If we assume that all OH/IR stars are evolved
from the progenitors of similar $M_{zams}$ and chemical abundance and the type E ones are the precursors
of the type A ones as we expected, it is reasonable to expect that both the luminosity and the mass
loss rate of the type E ones should be smaller than that of the type A ones. But in Fig. 4 of our paper,
many type E stars can reach high luminosity while their mass loss rates are always smaller than
that of the type A stars with the same luminosity. From our Fig. 1,2,6 and 7, we can also see that some type
E stars can evolve into luminous objects even if their period is shorter than type A ones.
This give us the hint that some of the two types of OH/IR stars may not evolve from the same population of
progenitors (with similar $M_{zams}$ and chemical abundance). If this is the right, we can safely infer
from the single $\dot M-P$ relation obeyed by the whole sample that the differences in $M_{zams}$ and 
chemical abundance of the progenitors have little effect in the $\dot M-P$ relation of OH/IR stars.

\cite{che01} have argued that there are probably many type E OH/IR
stars which would never evolve into type A ones, because they found that type A 
OH/IR stars are mainly located on the galactic plane while many type E stars
are located at high galactic latitude (see their Fig. 5). Here a similar
figure is given in Fig. 5a which shows a similar distribution of different
types of OH/IR stars. In addition, the more detailed position information of these 
stars are also shown in Fig. 5b and 5c. The distances from the stars to the 
galactic plane and from the projection of these stars on the galactic plane to 
the galactic center are plotted in Fig. 5b while the projection of these stars on 
the galactic plane is shown in Fig. 5c. It is indicated from Fig. 5 that the majority of 
type E sample stars are nearby sources while the other types distribute
over a larger area. There are three galactic arms near the sun:
the one at several dozen parsecs outside the solar circle is the Orion arm; the one at about 2 kpc
outside the solar circle is the Perseus arm and another at about 2 kpc inside the solar
circle is the Sagittarius arm. From Fig. 5c, it is reasonable to suppose that
many type A sample stars are located in the three galactic arms. Those
stars whose distance to the galactic center is smaller than about 5kpc must be
located inside the dense galactic HI cloud disk. And those type E stars are
located either in the inter-arms area on the galactic plane like the sun or outside of 
the galactic plane. From the space distribution of the sample stars, it is not difficult to
conclude that, in the whole sample, the type E and other type stars are probably
formed in different interstellar environment, hence their chemical
composition and evolution process can be different. This
probably is the original cause of the discrepancy between the $L_{*}-P$
and $\dot M-L_{*}$ relations of the type E and A stars. It should be noted that the
distance of many stars are not very accurate, so Fig. 5b,c are only valuable for reference.

The conclusions above are mainly based on the quantities derived by
modeling. But there exist many uncertainties in the modeling method.
The observed data are not very certain because of the variation of light, interstellar
extinction and emission. The distances to many OH/IR stars are not very accurate.
There are also many uncertainties in the existing radiative transfer modeling. For example,
the uncertain dust optical properties, the unknown dust grain physical status and density
distribution, etc. The spectrum of the center star should be calculated in more detail by
stellar photosphere models instead of using a blackbody spectrum. And many other
physical processes such as the dust formation process, the dust-gas interaction process
etc. need to be included in the model.
\begin{figure}[]
   \centering
	 \includegraphics[width=9cm]{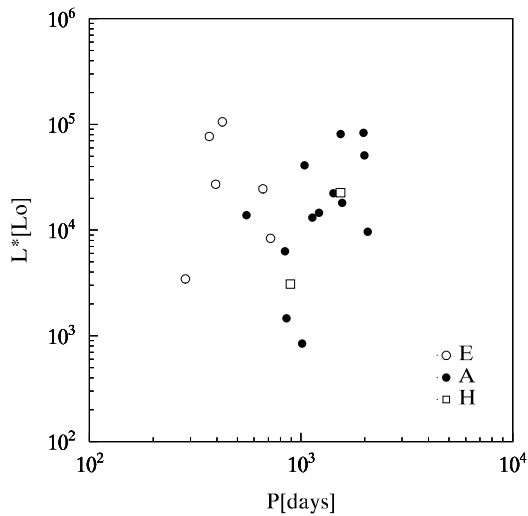}
   \caption{The $L_*-P$ relation for the phase-lag sample. The luminosity shown is
obtained by directly integrating the observed SEDs. }	
   \label{Fig6}
   \end{figure}
\begin{figure}[]
   \centering
	 \includegraphics[width=9cm]{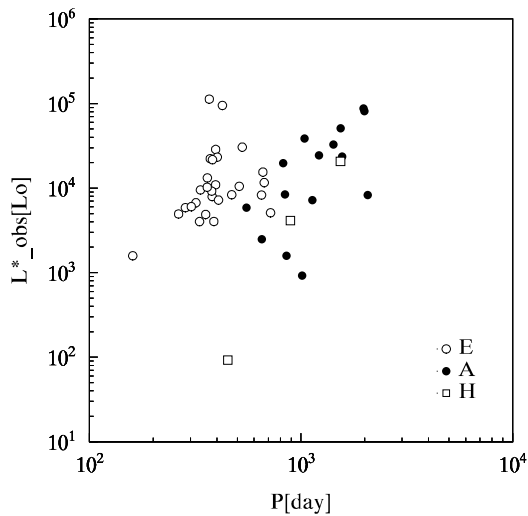}
   \caption{The $L_*-P$ relation for the whole sample. The luminosity shown is obtained by directly
integrating the observed SEDs. }	
   \label{Fig7}
   \end{figure}

Because of these drawbacks of the modeling method, it
is reasonable to question on the reliability of the derived data such as
luminosity and mass loss rate. In order to clarify the doubts about the derived
data, we also calculated the luminosity by directly integrating the
observed SEDs. When the spectrum is integrated, it is represented by the blackbody
spectrum of temperature of $200$ K (peak at about $25\mu$) in the wavelength range shorter
than the first NIR observed data point and longer than $100\mu$. 
The resulted $L_{*}-P$ relations are shown in Fig. 6 for the phase-lag sample and 
in Fig. 7 for the whole sample. In these two figures, the conclusion is basically the same.
This ensures in some degree the reliability of the data derived by modeling and 
the conclusions derived from them.

Another problem should also be noted: some of the sample stars, even some type E 
ones, have their luminosity very large (about $10^5L_{\sun}$); this is
unusual for OH/IR stars. In order to verify whether any calculation errors have
occurred, the luminosity data of some sample stars are compared with 
that from \cite{lan90} who had estimated luminosity
for some OH/IR stars with phase-lag distances in another way. The luminosity
data derived in this paper are found to be comparable with theirs. And there are 
also some stars in their paper reach a high luminosity near to $10^5L_{\sun}$. It is 
reasonable to suspect that these luminous stars may actually be supergiants. But further
inspection is needed for this subject.

\acknowledgements
We thank Prof. R. Szczerba who has kindly provided us the
radiative transfer code. We also have used in our work the online SIMBAD
database and the online IRAS Data Analysis Facility at the University of
Calgary. This work is supported by grants from National Natural Science
Foundation of China and Chinese Academy of Sciences.

\clearpage

\begin{table*}[]{}
\centering
\caption[]{References for NIR and distance data. }
\label{tbl-1}
\begin{tabular}{c@{ }r@{  }r@{  }r@{ }r@{  }r@{  }r@{  }r@{  }r@{}}
\hline
\noalign{\smallskip}
{IRAS\_name} &{NIR} & {dist.}   & {IRAS\_name} &{NIR}   & {dist.}   & {IRAS$_{name}$} &{NIR}   & {dist.}   \\
\noalign{\smallskip}
\hline
\noalign{\smallskip}
01037+1219   &21 &15 &14297$-$6010 &15   &13 &23412$-$1533 &5          &13  \\
01304+6211   &21 &15 &15193+3132   &22   &14 &23425+4338   &22         &16  \\
03293+6010   &18 &3  &16474$-$4418 &15   &13 &15255+1944   &21         &1   \\
03336$-$7636 &5  &16 &16494$-$4327 &15   &13 &18182$-$1504 &7          &9   \\
03507+1115   &5  &9  &17073$-$4225 &15   &13 &18257$-$1000 &2,4,7      &19  \\
04140$-$8158 &5  &16 &17073$-$3955 &15   &13 &18257$-$1052 &2,11,17    &19  \\
05098$-$6422 &5  &16 &17271$-$3425 &15   &13 &18276$-$1431 &15         &9,1 \\
05528+2010   &21 &1  &17317$-$3331 &15   &13 &18348$-$0526 &21         &19  \\
06500+0829   &12 &16 &17360$-$3012 &15   &13 &18431$-$0403 &8,15       &19  \\
07209$-$2540 &5  &13 &17540$-$1919 &12   &14 &18432$-$0149 &8,15,17    &9   \\
07585$-$1242 &5  &16 &18135$-$1456 &15   &9  &18445$-$0238 &4,11,15    &9   \\
09425+3444   &22 &16 &18437$-$0643 &15   &9  &18460$-$0254 &2,4,11,15  &9   \\
09429$-$2148 &5  &3  &19017+0608   &2,18 &3  &18488$-$0107 &2,8,10,11  &19  \\
10189$-$3432 &5  &16 &19059$-$2219 &5    &13 &18498$-$0017 &4,11,12,17 &19  \\
10287$-$5733 &15 &13 &20000+4954   &18   &9  &18549+0208   &8          &19  \\
10580$-$1803 &5  &16 &20077$-$0625 &5    &13 &18560+0638   &21         &19  \\
12310$-$6233 &15 &13 &20259$-$4035 &20   &14 &19039+0809   &21         &9   \\
13157$-$6421 &15 &13 &21206$-$4054 &20   &14 &19192+0922   &6          &19  \\
14086$-$2839 &5  &14 &22516+0838   &5    &13 &19550$-$0201 &5          &1   \\
14247+0454   &5  &9  &23041+1016   &5    &14 &22177+5936   &21         &9   \\
              \noalign{\smallskip}               
\hline 
\end{tabular}

\begin{list}{}{}
\item[] 
Reference list:
(1)  \cite{bau83}; (2)  \cite{bau85}; (3)  \cite{bow83};
(4)  \cite{eva84}; (5)  \cite{fou92}; (6)  \cite{fix84};
(7)  \cite{gar97}; (8)  \cite{geh85}; (9)  \cite{her86};
(10) \cite{kwo87}; (11) \cite{jon83}; (12) \cite{lep95};
(13) \cite{sid96}; (14) \cite{ngu79}; (15) \cite{nym93};
(16) \cite{ort96}; (17) \cite{sch76}; (18) \cite{sun98};
(19) \cite{lan90}; (20) \cite{whi94}; (21) \cite{xio94};
(22) query from SIMBAD database
\end{list}

\end{table*}

\begin{table*}[]{}
\centering
\setcounter{table}{1}
\caption[]{Some parameters of the stars with phase-lag distances. }
\label{LineOHpump}
\begin{tabular}{c@{ }r@{     } r@{      }r@{     }r@{      }r@{      }r@{      }}
\hline
\noalign{\smallskip}
IRAS\_name  & period  & $V_e$   & D & $L_*$   & $\dot M$   & LRS type\\
                     & [day]    & $[km/s]$ &  [kpc] &       [$10^3L_{\sun}$]   &          [$M_{\sun}$/yr]   &         \\
\noalign{\smallskip}
\hline
\noalign{\smallskip}
01037+1219......   &660.0  & 18.2 & 0.74  & 24.56  & 0.96  & E\\ 
05528+2010......   &368.3  & 2.7  & 1.20  & 76.86  & 0.11  & E\\ 
15255+1944......   &425.1  & 7.5  & 3.10  & 105.71 & 0.58  & E\\ 
19039+0809......   &284.2  & 6.6  & 0.29  & 3.46   & 0.02  & E\\ 
18431$-$0403...... &717.0  & 17.5 & 1.45  & 8.36   & 0.24  & E\\ 
19550$-$0201...... &394.8  & 5.4  & 0.85  & 27.20  & 0.06  & E\\ 
01304+6211......   &1995.0 & 11.0 & 2.90  & 50.91  & 6.86  & A\\ 
18348$-$0526...... &1566.0 & 14.0 & 1.44  & 18.09  & 4.50  & A\\
18560+0638......   &1424.0 & 16.1 & 2.04  & 22.31  & 4.93  & A\\
22177+5936......   &1215.0 & 14.5 & 2.38  & 14.54  & 5.58  & A\\
18182$-$1504...... &1011.7 & 20.5 & 0.55  & 0.84   & 0.78  & A\\
18257$-$1052...... &1129.5 & 18.2 & 5.06  & 13.09  & 2.04  & A\\
18257$-$1000...... &1975.2 & 19.0 & 8.51  & 83.31  & 18.99 & A\\
18276$-$1431...... &890.0  & 12.0 & 2.08  & 3.09   & 0.60  & H\\
18432$-$0149...... &1038.6 & 17.5 & 7.70  & 40.99  & 8.26  & A\\
18445$-$0238...... &853.0  & 17.0 & 1.09  & 1.46   & 0.58  & A\\
18460$-$0254...... &2064.4 & 19.5 & 1.77  & 9.67   & 3.59  & A\\
18488$-$0107...... &1540.2 & 20.6 & 11.87 & 81.31  & 16.32 & A\\
18498$-$0017...... &1536.3 & 15.9 & 5.02  & 22.58  & 1.58  & H\\
18549+0208......   &840.0  & 13.6 & 4.21  & 6.30   & 0.91  & A\\
19192+0922......   &552.0  & 16.1 & 1.13  & 13.82  & 0.52  & A\\
              \noalign{\smallskip}               
\hline 
\end{tabular}
\end{table*}


\begin{thebibliography}{}
\bibitem[Baud and Habing(1983)]{bau83} Baud, B., Habing, H. J.  1983, \aap, 127, 73
\bibitem[Baud et al.(1985)]{bau85} Baud, B., Sargent, A. I., Werner, M. W., Bentley, A. F.  1985, \apj, 292, 628
\bibitem[Bowers et al.(1983)]{bow83} Bowers, P. F., Johnston, K. J., Spencer, J. H.  1983, \apj, 274, 733
\bibitem[Blommaert et al.(1998)]{blo98} Blommaert, J. A. D. L., van der Veen, W. E. C. J., van Langevelde, H. J., Habing, H. J., Sjouwerman, L. O., 1998, \aap, 329, 991
\bibitem[Cardelli et al.(1989)]{car89} Cardelli, J. A., Clayton, G. C., Mathis, J. S.  1989, \apj, 345, 245
\bibitem[Chen et al.(2001)]{che01} Chen, P. S., Szczerba, R., Kwok, S., Volk, K.  2001, \aap, (to be published)
\bibitem[Drain(1985)]{dra85} Drain, B. T.  1985, \apjs, 57, 587 
\bibitem[Drain and Lee(1984)]{dra84} Drain, B. T., Lee, H. M.  1984, \apj, 285, 89
\bibitem[Evans and Bechwith(1984)]{eva84} Evans II, N. J., Beckwith, S.  1977, \apj, 217, 729 
\bibitem[Fix and Mutel(1984)]{fix84} Fix, J. D., Mutel, R. L.  1984, \aj, 89, 406
\bibitem[Fouque et al.(1992)]{fou92} Fouque, P., Le Bertre, T., Epchtein, N., Guglielmo, F., Kerschbaum, F.  1992, \aaps, 93, 151 
\bibitem[Garcia-Lario et al.(1997)]{gar97} Garcia-Lario, P., Manchado, A., Pych, W., Pottasch, S. R.  1997, \aaps, 126, 479
\bibitem[Gehrz et al.(1985)]{geh85} Gehrz, R. D., Hackwell, J. A., Jones, T. W.  1985, \apj, 290, 296
\bibitem[Herman et al.(1986)]{her86} Herman, J., Burger, J. H., Penninx, W. H.  1986, \aap, 167, 247
\bibitem[Jones et al.(1983)]{jon83} Jones, T. J., Hyland, A. R., Gatley, I.  1983, \apj, 273, 660
\bibitem[Kwok et al.(1987)]{kwo87} Kwok, S., Hrivnak, B. J., Boreiko, R. T.  1987, \apj, 321, 975 
\bibitem[Le Bertre(1993)]{ber93} Le Bertre, T.  1993, \aaps, 97, 729 
\bibitem[Lepine et al.(1995)]{lep95} Lepine J. R. D., Ortiz, R., Epchtein, N.  1995, \aap, 299, 453 
\bibitem[Le Sidaner and Le Bertre(1996)]{sid96} Le Sidaner, P., Le Bertre, T.  1996, \aap, 314, 896 
\bibitem[Mathis et al.(1977)]{mat77} Mathis, J. S., Rumpl, w., Nordsieck, K. H.  1977, \apj, 217, 425
\bibitem[Nguyen-Q-Rie et al.(1979)]{ngu79} Nguyen-Q-Rieu, Mr., Laury-Micoulaut, C., Winnberg, A., Schultz, G. V.  1979, \aap, 75, 351
\bibitem[Nyman et al.(1993)]{nym93} Nyman, L.-A., Hall, P. J., Le Bertre, T.  1993, \aap, 280, 551
\bibitem[Ortiz and Maciel(1996)]{ort96} Ortiz, R., Maciel, W. J.  1996, \aap, 313, 180 
\bibitem[Schultz et al.(1976)]{sch76}Schultz, G. V., Kreysa, E., Sherwood, W. A.  1976, \aap, 50, 171
\bibitem[Szczerba et al.(1997)]{szc97} Szczerba, R., Omont, A., Volk, K., Cox, P., Kwok, S.  1997, \aap, 317, 859
\bibitem[Sun and Zhang(1998)]{sun98} Sun J., Zhang H. Y.  1998, ChA\&A,22, 442 
\bibitem[van Herk(1965)]{her65} van Herk, G.  1965, Bull. Astron. Inst. Neth. 18, 71 
\bibitem[van Langevelde et al.(1990)]{lan90} van Langevelde, H. J., van der Heiden, R., van Schooneveld, C.  1990, \aap, 239, 193
\bibitem[Whitelock et al.(1991)]{whi91} Whitelock, P., Feast, M., Catchpole, R., 1991, \mnras,248, 276
\bibitem[Whitelock et al.(1994)]{whi94} Whitelock, P., Menzies, J., Feast, M., Marang, F., Carter, B., Roberts, G., Catchpole, R., 
     Chapman, J.  1994, \mnras,267, 711
\bibitem[Xiong et al.(1994)]{xio94} Xiong, G. Z., Chen, P. S., Gao, H.  1994, \aaps, 108, 1 
\end{thebibliography}
\end{document}